\documentclass[11pt]{article}
\usepackage{graphicx}
\usepackage{latexsym}
\usepackage{setspace} 
\def\tr{{\rm tr}}
\def\diag{{\rm diag}\,}

\def\C{{\bf C}}
\def\Z{{\bf Z}}

\def\CD{{\cal D}}

\def\CN{{\cal N}}
\def\CO{{\cal O}}

\def\centeron#1#2{{\setbox0=\hbox{#1}\setbox1=\hbox{#2}\ifdim
   \wd1>\wd0\kern.48\wd1\kern-.48\wd0\fi
   \copy0\kern-.48\wd0\kern-.48\wd1\copy1\ifdim\wd0>\wd1
   \kern.48\wd0\kern-.48\wd1\fi}}

\newcommand{\beq}{\begin{equation}}
\newcommand{\eeq}{\end{equation}}
\newcommand{\bea}{\begin{eqnarray}}
\newcommand{\eea}{\end{eqnarray}}
\newcommand{\ba}{\begin{array}}
\newcommand{\ea}{\end{array}}

\newcommand{\p}{\partial}
\newcommand{\nn}{\nonumber}
\newcommand{\bp}{\bar{\partial}}

\newcommand{\half}{\frac{1}{2}}

\begin{document}

\hskip6cm

\centerline{\LARGE \bf   Eight Fermion Terms  in the Effective Action }
\vskip0.5cm \centerline{\LARGE \bf  of the ABJM Model}

\vskip2cm

\vskip2cm

\centerline{Jong-Hyun Baek$^*$,  Seungjoon Hyun$^{*\dagger}$ and  Sang-Heon Yi$^*$}

\hskip1.7cm

\centerline{\it ${}^*$Department of Physics, College of Science, Yonsei University, Seoul 120-749, Korea}
\centerline{\it ${}^\dagger$Korea Institue for Advanced Study, Seoul 130-722, Korea}

\vskip3cm

\centerline{\bf Abstract}
 We study eight fermion terms in the effective action of the ABJM model. We show the non-renormalization of $v^2$ terms. After classifying all the possible eight fermion structures, we show that $\CN=6$ supersymmetry determines all these terms completely up to an overall constant. This confirms the one loop non-renormalization of $v^4$ terms.
\thispagestyle{empty}
\renewcommand{\thefootnote}{\arabic{footnote}}
\setcounter{footnote}{0}
\newpage

%
\section{Introduction and Conclusion}
Recently there have been much interests in the $\CN=6$ superconformal $U(N)\times U(N)$  Chern-Simons-matter theory (ABJM Model) \cite{Aharony:2008ug} which is conjectured to be dual to M-theory on $AdS_4\times S^7/\Z_k$. Despite many interesting results on the model, 
yet  a deeper understanding of the model  and more supporting evidences on the duality are still needed. 
One basic test for the duality is to study the effective action corresponding to the membrane scattering in M-theory~\cite{Verlinde:2008di}\cite{Baek:2008ws}\cite{Berenstein:2008dc}\cite{Hosomichi:2008ip}. In field theory side,  it corresponds to  the effective action for  the symmetry breaking, in the simplest context, of $U(2)  \times U(2)$ to
$U(1)\times U(1)\times U(1)\times U(1)$ by giving the vacuum expectation values to the scalar fields as $\langle Y^A\rangle=\diag(0, b^A)$. In the dual gravity description,  it corresponds to the motion of a probe brane at $b^A$  in the background of   $AdS_4\times S^7/\Z_k$.

One may expect that the $\CN=6$ superconformal symmetry strongly restricts the possible form of the effective action, in particular, the lower order terms in derivative expansions.  It has been known that sixteen supersymmetries play an essential role in the analogous study in the matrix model and super Yang-Mills theories~\cite{Dine:1997nq}\cite{Paban:1998ea}\cite{Paban:1998qy}\cite{Paban:1998mp}\cite{Hyun:1999hf}. It was found that the $v^2$ terms and their superpartners do not receive any quantum corrections while the $v^4$ terms and their superpartners are one-loop exact modulo non-perturbative corrections.

One convenient way to organize terms in the derivative expansions of the effective action is to assign the appropriate weight to the fields~\cite{Dine:1997nq}. The scalar  and spinor fields are assigned to have weight  zero and one half, respectively, and the  derivatives have one. The kinetic terms in the classical action has weight two. Next nontrivial terms in the effective action have weight four among which they are related by supersymmetry.  In this context, the most crucial terms are those with the largest number of fermions, i.e. eight fermion terms, as they are typically determined among themselves by supersymmetry.  
Once they are determined, we can use the supersymmetry to determine all the other weight four terms, in particular the $v^4$ terms.     

 In this short note we study the effective action of the ABJM model  involving eight fermion terms, which  are superpartners of the $|\partial b|^4$ terms, by using the supersymmetry. First of all we show the non-renormalization of $|\partial b|^2$ terms. This implies that there are no corrections to the supercharges up to the order we are interested in. 
The eight fermion terms generically contain scalar fields. Since there  are no corrections in supercharges, the supersymmetry transformations acting on these scalar fields in eight fermion terms are the only source for nine fermion terms and therefore they should vanish by themselves. In this way we determine the eight fermion terms modulo an overall constant. This implies that the  $|\partial b|^4$ terms and their superpartners are completely determined by supersymmetry. In fact since they come from one loop, our results naturally imply one-loop non-renormalization of  the $|\partial b|^4$ terms.

\section{Non-renormalization of $v^2$ Terms}
We are interested in the effective action of the slowly moving probe brane. The target space coordinates of the branes are described by the diagonal components of the scalar fields. We put the source branes at the origin of $\C^4/\Z_k$ and the probe brane at $b^A$.  The superpartners of these scalars, $b^A$, are denoted as $\chi_A$. After integrating out the off-diagonal components of scalar, spinor and gauge fields, the only relevant fields remained are diagonal components of those fields. One combination of the remaining abelian gauge fields is decoupled from all matter fields and thus can be integrated out trivially. This gives the constraints on the other combination of the gauge fields to be pure gauge, and thus to be zero. The only remaining gauge symmetry is the global discrete one, $\Z_k$, identifying  $b^A \sim e^{2\pi i/k}b^A$ and $\chi_A\sim e^{2\pi i/k}\chi_A$.
Therefore the effective action becomes a functional of $b^A$ and $\chi_A$. 
 
The  tree level supersymmetry transformations of these fields become
\bea \delta b^A = i\epsilon^I\gamma^{AB}_{I}\chi_B\,, \qquad
     \delta \chi_{A\, \alpha} = ({\gamma}^{\mu}{\epsilon}^I)_{\alpha}\bar{\gamma}_{I\, AB}\, \p_{\mu} b^B\,,
\eea
where $\gamma_{\mu}$ and $\gamma_{I}$,  $\bar{\gamma}_{I}$ denote the $SO(2,1)$ and the $SO(6)$ gamma matrices, respectively. Various properties on these gamma matrices are summarized in Appendix. We always contract spinor indices from northwest to southeast, $\psi \chi \equiv 
\psi^{\alpha}\chi_{\alpha}$, which gives 
$\psi \chi =\chi\psi$  and $\psi\gamma^{\mu}\chi=-  \chi\gamma^{\mu}\psi$. 

In general there could be loop corrections to the supersymmetry transformations. The generic form of the corrections can also be organized by weights.  The lowest order terms in the effective action have weight two, which is the same as the classical action. Therefore the correction in the supersymmetry transformations at this order, if any, should have the same weight as the tree level supersymmetry transformations and thus generically given by       
\bea
\delta b^A = i\epsilon^I\gamma^{AB}_{I}\chi_B\,, \qquad
\delta \chi_{A\,\alpha} = {({\gamma}^{\mu} {\epsilon}^I)}_{\alpha}\bar{\gamma}_{I\, AB}\,\p_{\mu} b^B + {(M_{I\, A} \epsilon^I)}_{\alpha}\,,
\eea
where $M_{I}$ contain fermion bilinears. The supersymmetry algebra should remain closed under these modified transformations and thus from
\bea
[\delta_1, \delta_2]b^A &=& i\gamma^{AB}_I\left\{\epsilon^I_2({\gamma}^{\mu}{\epsilon}^J_1\bar{\gamma}_{J\, BC}\, \p_{\mu}b^C + M_{J\, B}\epsilon^J_1)-\epsilon^I_1({\gamma}^{\mu}{\epsilon}^J_2\bar{\gamma}_{J\, BC}\, \p_{\mu}b^C + M_{J\,B}\epsilon^J_2)\right\} \nn\\ &=& 2i(\epsilon^I_1{\gamma}^{\mu}{\epsilon} _{2\, I})\,\p_{\mu}b^A + i\gamma^{AB}_I(\epsilon^I_2 M_{J\,B}\epsilon^J_1 - \epsilon^I_1 M_{J\,B}\epsilon^J_2)\,,
\eea
 one can see that the second term should vanish.  As shown in below, this condition demands that $M_J$ be zero and thus the supercharges do not get any corrections at this order. This guarantees that the $|\partial b|^2$ terms do not get any quantum correction. 

Proof: The condition which 
$M_J$ should satisfy is
\beq\label{first}
 \gamma_J^{AB} {M_{IB}}^{\beta\alpha} + \gamma_I^{AB} {M_{JB}}^{\alpha\beta} =0\,. \label{condition}
\eeq
Multiplying by ${\bar{\gamma}}^I_{CA}$, the equation becomes
\beq  
{\bar{\gamma}}^I_{CA}\gamma_J^{AB}{M_{IB}}^{\beta\alpha} - 6{M_{JC}}^{\alpha\beta}=0 \,. \label{cond1}
\eeq
By symmetrizing the spinor indices and using the $SO(6)$ Clifford algebra this becomes
\bea  -{\bar{\gamma}}_{JCA}\gamma^{IAB}(M_{IB}^{\,\alpha\beta}+ M_{IB}^{\, \beta\alpha})- 8(M_{JC}^{\,\alpha\beta} + M_{JC}^{\, \beta\alpha})=0\,. \nn
\eea
By multiplying $\delta^{IJ}$ in the~Eq.(\ref{condition}) and plugging in the above equation, one can see that $M^I_A$ are antisymmetric, 
$ M^{I}_{~ A\,\beta\alpha} = -M^{I}_{~A\, \alpha\beta}$.
Therefore, the Eq.(\ref{cond1}) becomes 
$  {\bar{\gamma}}^J_{CA}\gamma_I^{AB} M^{I\,\alpha \beta}_B -4 M^{J\,\alpha\beta}_C=0$.
By multiplying $\gamma_{J}$, we obtain that $\gamma_IM^{I}=0$, which leads to
$  M^{I}_{~A\, \alpha\beta} =0$. $\clubsuit$

\section{Fermion Bilinears}
In this section the possible fermion bilinears are classified.
Note that $\chi^{\dag A}$ and $\chi_A$ transform as $(\bar{\bf 2},{\bf 4})$ and $({\bf 2},\bar{\bf 4})$ under the three-dimensional Lorentz symmetry and the $R$-symmetry $SO(2,1)\times SO(6)_R$, respectively. We can classify the fermion bilinears according to the irreducible representations of  $SO(2,1)\times SO(6)$ as follows.
\begin{itemize}
\item \underline{$\chi^{\dag}\chi^{\dag}$}: \qquad \qquad \qquad
$  (\bar{\bf 2}, {\bf 4})\times(\bar{\bf 2},{\bf 4})
 = ({\bf 1},{\bf 6}) + ({\bf 1},{\bf 10}) + ({\bf 3},{\bf 6}) + ({\bf 3},{\bf 10}). \\
$
Each irreducible representation corresponds to the following fermion bilinear form:
\bea
 \chi^{\dag A}\bar{\gamma}^I_{AB}\chi^{\dag B}\,,  \qquad
 \chi^{\dag A}(\bar{\gamma}^{IJK})_{AB}\chi^{\dag B}\,, \qquad
\chi^{\dag A}\gamma^{\mu}\bar{\gamma}^I_{AB}\chi^{\dag B}\,, \qquad
 \chi^{\dag A}\gamma^{\mu}(\bar{\gamma}^{IJK})_{AB}\chi^{\dag B}\,.
\eea
Among these, the first and fourth terms identically vanish by the (anti-)symmetry of $SO(6)$ gamma matrices.
\item 
\underline{$\chi\chi$}: \qquad \qquad \qquad
$ 
({\bf 2}, \bar{\bf 4})\times({\bf 2},\bar{\bf 4}) = ({\bf 1},{\bf 6}) + ({\bf 1},{\bf 10}) + ({\bf 3},{\bf 6}) + ({\bf 3},{\bf 10})
$ \\
Each irreducible representation corresponds to the following fermion bilinear form:
\bea
\chi_A \gamma_I^{AB}\chi_B\,, \qquad
 \chi_A(\gamma_{IJK})^{AB}\chi_B\,, \qquad
\chi_A\gamma^{\mu}\gamma_I^{AB}\chi_B\,, \qquad \chi_A\gamma^{\mu}(\gamma_{IJK})^{AB}\chi_B\,,
\eea
where the first and fourth terms vanish by the same reason as above.
\item 
\underline{$\chi^{\dag}\chi$}: \qquad \qquad \qquad
$ ({\bf \bar{2}}, {\bf 4})\times({\bf 2},{\bf \bar{4}}) = ({\bf 1},{\bf 1}) + ({\bf 1},{\bf 15}) + ({\bf 3},{\bf 1}) + ({\bf 3},{\bf 15})\,.
$ \\
Each irreducible representation corresponds to the following fermion bilinear form:
\bea
\chi^{\dag A}\chi_A\,,\qquad
\chi^{\dag A}(\bar{\gamma}^{IJ})_A^{~~B}\chi_B\,, \qquad
\chi^{\dag A}\gamma^{\mu}\chi_A\,, \qquad
 \chi^{\dag A}\gamma^{\mu}(\bar{\gamma}^{IJ})_A^{~~B}\chi_B\,.
\eea
\end{itemize}

These fermion bilinears are the basic building blocks in the effective action.
All the eight fermion terms which appear as the superpartners of $|\partial b|^4$ should come from the combinations of the above fermion bilinears.

\section{Structures of The Eight Fermion Terms}

Eight fermion terms in the effective action should be a singlet under $SO(2,1)\times SO(6)$. In particular, this implies that all the $SO(6)$ vector indices should be contracted. Since they appear only through $SO(6)$ gamma matrices, we can use the $SO(6)$ gamma matrix identities in Appendix to have expressions involving $SU(4)$  indices only. 
Similarly, $SO(2,1)$ vector indices appear only through the derivatives $\partial_\mu b^A$ and  $SO(2,1)$ gamma matrices. Since eight fermion terms do not contain any derivative, we can use  the three dimensional Fierz identity 
to obtain expressions involving $SO(2,1)$ spinor indices only.

The requirements of being $SU(4)$ and gauge singlet  strongly restrict the possible form of eight fermion terms. Therefore the possible eight fermion terms, denoted collectively as $f\, \chi^8$,  can be written as 
\bea
f\,\chi^8 &=& f^{l_0}_{0}~ T_{l_0} + f^{l_2}_{2}~ b^{\dag}_{A}b^{B}\, T^{ A}_{l_2\, B} + f^{l_4}_{4}~ b^{\dag}_{A}b^{\dag}_{C}b^{B}b^{D}\, T^{AC}_{l_4\,~BD}   \nn\\
&&  {} + f^{l_6}_{6}~ b^{\dag}_{A}b^{\dag}_{C}b^{\dagger}_{E}b^{B}b^{D}b^{F}~ T^{ACE}_{l_6\,~~BDF}+ f^{l_8}_{8}~ b^{\dag}_{A}b^{\dag}_{C}b^{\dagger}_{E}b^{\dagger}_{G}b^{B}b^{D}b^{F}b^{H}~ T^{ACEG}_{l_8\,~~BDFH}\,,
\eea
where $T$'s are the possible eight fermion structures. By the symmetry of the given configurations, the coefficients $f^{l_i}_{n}$ are functions of a variable
$
r\equiv |b|=\sqrt{b^Ab^{\dagger}_A} $ only. 

One may notice that,  because of bosonic $b$ and $b^{\dagger}$ factors, $T$'s should be symmetrized among upper/lower indices. Because of anti-commuting nature of  fermions, one can easily show that 
\beq
 T^{(ACE)}_{~~~(BDF)} =0\,, \qquad \qquad   T^{(ACEG)}_{~~~~~(BDFH)} =0\,,
\eeq
where $(ACE)$ denotes the total symmetrization of $A,C,E$.

Now, let us classify all the possible eight fermion structrues for $T_{l_0}$, $T^{A}_{l_2\, B}$ and $T^{AB}_{l_4~ CD}$, which correspond to terms containing zero, two and four scalars, respectively. The Fierz identity, 
\beq
(\chi^{\dagger A}\cdot\chi^{\dagger C})(\chi_B\cdot\chi_D) = -(\chi^{\dagger A}\cdot\chi_B)(\chi^{\dagger C}\cdot\chi_D)-(\chi^{\dagger A}\cdot\chi_D)(\chi^{\dagger C}\cdot\chi_B)\,,
\eeq
can be used to replace both $\chi^{\dagger} \chi^{\dagger}$ and $\chi \chi$ contractions by $\chi^{\dagger} \chi$ contractions and vice versa. Here $\cdot$ denotes the contraction of spinor indices.

Firstly, it is clear from the above Fierz identity that  there are only two independent structures in eight fermions with four scalars, which are given by
\bea
T^{AC}_{1~\, BD} &=& T^{(AC)}_{1~\, (BD)}=(\chi^{\dagger A}\cdot\chi^{\dagger C})(\chi_{B}\cdot\chi_{D})(\chi^{\dagger E}\cdot\chi_{E})(\chi^{\dagger F}\cdot\chi_{F}) \,,\nn \\
T^{AC}_{2~\, BD} &=&T^{(AC)}_{2~\, (BD)} = (\chi^{\dagger A}\cdot\chi^{\dagger C})(\chi_{B}\cdot\chi_{D})(\chi^{\dagger E}\cdot\chi_{F})(\chi^{\dagger F}\cdot\chi_{E})\,. \nn
\eea

It is a bit more complicated to find the independent structures with two scalars. Apparently, there are seven possible structures, 
\bea
N^A_{1\,B} &=& (\chi^{\dagger A}\cdot\chi_B) (\chi^{\dagger C}\cdot\chi_{C})^3\,, \nn\\
N^A_{2\, B}  &=& (\chi^{\dagger A}\cdot\chi_C) (\chi^{\dagger C}\cdot \chi_B) (\chi^{\dagger D}\cdot\chi_{D})^2\,,  \nn\\
N^A_{3\, B} &=& (\chi^{\dagger A}\cdot\chi_B)(\chi^{\dagger C}\cdot\chi_{C}) (\chi^{\dagger E}\cdot\chi_F)(\chi^{\dagger F}\cdot\chi_E)\,, \nn\\
N^A_{4\, B}  &=& (\chi^{\dagger A}\cdot\chi_C) (\chi^{\dagger C}\cdot \chi_B) (\chi^{\dagger E}\cdot\chi_F)(\chi^{\dagger F}\cdot\chi_E)\,, \\
N^A_{5\, B}  &=& (\chi^{\dagger A}\cdot\chi_B) (\chi^{\dagger C}\cdot\chi_D)(\chi^{\dagger D}\cdot\chi_E)(\chi^{\dagger E}\cdot\chi_C)\,, \nn\\
N^A_{6\, B}  &=& (\chi^{\dagger A}\cdot\chi_C)(\chi^{\dagger C}\cdot\chi_D)(\chi^{\dagger D}\cdot\chi_B) (\chi^{\dagger E}\cdot\chi_{E})\,, \nn\\
N^A_{7\, B}  &=& (\chi^{\dagger A}\cdot\chi_C)(\chi^{\dagger C}\cdot\chi_D)(\chi^{\dagger D}\cdot\chi_E) (\chi^{\dagger E}\cdot\chi_B)\,. \nn
\eea
It can be shown that they are related by three equations as
\beq
 N_2+N_6  = N_3+N_5 = -\half(N_1+N_3)\,, \qquad N_6 + N_7 = -\half (N_2+N_4)\,.
\eeq
Therefore, there are four independent structures with two scalars, which may be chosen as
\bea
T^{A}_{1\, B} &\equiv&   -{(N_1+N_2)}^A_{\, B} = T^{AC}_{1~ BC}\,,\qquad 
T^{A}_{2\, B} \equiv -{(N_3+N_4)}^A_{\, B} = T^{AC}_{2~ BC}\,, \qquad T^{A}_{3\, B} \equiv N^{A}_{3\, B}\,, \nn  \\
T^{A}_{4\, B} &\equiv& -{(N_3+N_6)}^A_{\, B}=(\chi^{\dagger A}\cdot\chi^{\dagger C})(\chi^{\dagger D}\cdot\chi_{C})(\chi_{D}\cdot\chi_{B})(\chi^{\dagger E}\cdot\chi_{E})\,.
\eea
One may note that the contraction of two indices in eight fermion structures with four scalars gives rise to two independent ones with two scalars.

Similarly, there are five possible  structures without scalars,
\bea
 M_1 &\equiv&   (\chi^{\dagger A}\cdot\chi_{A})^4 = N^A_{1\, A}\,,  \nn\\
 M_2 &\equiv&  (\chi^{\dagger A}\cdot\chi_{A})^2(\chi^{\dagger C}\cdot\chi_D)(\chi^{\dagger D}\cdot\chi_C) = N^A_{2\, A}=N^A_{3\, A}\,,\nn\\
M_3 &\equiv&    (\chi^{\dagger A}\cdot\chi_C) (\chi^{\dagger C}\cdot \chi_A) (\chi^{\dagger E}\cdot\chi_F)(\chi^{\dagger F}\cdot\chi_E)= N^A_{4\, A}\,, \nn\\
 M_4 &\equiv&   (\chi^{\dagger A}\cdot\chi_{A}) (\chi^{\dagger C}\cdot\chi_D)(\chi^{\dagger D}\cdot\chi_E)(\chi^{\dagger E}\cdot\chi_C)=N^A_{5\, A}= N^A_{6\, A}\,, \\
 M_5 &\equiv&    (\chi^{\dagger A}\cdot\chi_C)(\chi^{\dagger C}\cdot\chi_D)(\chi^{\dagger D}\cdot\chi_E) (\chi^{\dagger E}\cdot\chi_A) = N^A_{7\, A}\,, \nn 
 \eea
which are related by two equations,
\bea
 M_4 = -\frac{1}{2}M_1 - \frac{3}{2}M_2~, \qquad
 M_5 &=& \frac{1}{2}M_1 + M_2 - \frac{1}{2}M_3 ~.
 \eea
We choose three independent structures as 
\bea T_1 \equiv  -(M_1+M_2)  = T^{A}_{1~ A}= -2T^{A}_{4~ A}\,, \qquad
      T_2 \equiv -(M_2+M_3)  = T^{A}_{2~ A}\,,   \qquad
    T_3 \equiv M_2= T^{A}_{3~ A}\,.  ~~~  \eea
In summary, the effective action can have, at most, nine independent eight fermion structures. In next section, by using the supersymmetry, we show that some of these terms can not appear while all the remaining terms should be related.

\section{Determination of The Eight Fermion Terms}

In general, the tree level supercharges of the classical action get quantum corrections which may be organized by weights. 
Generically, the effective action can be expanded in the increasing order of the weights, such that the weights of consecutive terms differ by two. Accordingly, the corrections in supercharges should be ordered in the same way. The tree level supersymmetry transformations acting on scalars in eight fermion terms give rise to nine fermion terms.  On the other hand, in section 2, we showed that the supercharges do not get any corrections in the leading order and as a result the moduli space is flat. This implies that the nine fermion terms which arise from the supersymmetry variations of eight fermion terms should vanish by themselves.
Therefore we require that
\bea
\delta_{boson}\left(f\,\chi^8\right)   \equiv  \epsilon^{I\,\alpha}F_{I\,\alpha}\chi^8 = {\epsilon}^{I \alpha}\left[(\p_Af)(\gamma^{IAB}\chi_{B\alpha}) + (\bp^A f)(\bar{\gamma}^{I}_{AB}{\chi}^{\dag B}_{\alpha})\right]\chi^8 
= 0\,.
\eea

In order to solve these equations, it is convenient to apply the operator $\CO^{I\, \alpha}_1= \bar{\gamma}^{I}_{CD}\bar{\p}^{D}\frac{\p}{\p \chi_{C\alpha}}$ to $F_{I\alpha}\chi^8$, 
 from which we obtain the simple equation $ \CO (f\chi^8 )=0 $ with
\beq \CO = -12 \bp^{A}\p_{A}+ 2\bp^{A}\p_{A}~\chi_{ B\, \alpha}\frac{\p}{\p\chi_{B\, \alpha}} -2\p_{B} \bp^{A}~\chi_{A\, \alpha}\frac{\p}{\p\chi_{B\, \alpha}}\,. \eeq
To simplify further, we  introduce a fermion number operator for $\chi$ 
\beq \CO_{\chi} \equiv \chi_{A\, \alpha}\frac{\p}{\p \chi_{A\, \alpha}}\,,
\eeq
which gives 
$ \CO_{\chi}(f\chi^8 ) = 4(f\chi^8 )$. 
This operator has the following commutation relation with the operator $\CO$ :
\[
 [\CO_{\chi}, \CO] =  -2 \p_{B}\bp^{A}\chi_{A\, \alpha}\frac{\p}{\p \chi_{B\, \alpha}}\,, 
\]
which leads, along with  $\CO (f\chi^8) =0$, to  the relation
\begin{equation}
 \p_{B}\bp^{A}\chi_{A\, \alpha}\frac{\p}{\p \chi_{B\, \alpha}} (f\chi^8) =0 \,. \label{other}
\end{equation}
As a result, eight fermion terms should satisfy
\beq  \Delta (f\chi^8) =0\,. \eeq
where $\Delta= 4 \partial_A\bar{\partial}^A$ is an eight-dimensional Laplacian. These two equations, supplemented with simple dimensional counting, are enough to determine the eight fermion terms completely(up to an overall constant).

By expanding this eight dimensional Laplace equation in independent eight fermion structures, one obtains nine differential equations. Among these, the differential equations from the coefficient of $T_{3B}^A$ and  $T_{4B}^A$ are given by
\begin{eqnarray}
   \Big(\frac{d^2}{dr^2} + \frac{11}{r}\frac{d}{dr}\Big)f^{l}_2  =0~, \qquad  l=3,4.
   \end{eqnarray}
The nontrivial solution is of the form $f^{l}_2  \sim r^{-10}$, which can not appear by dimensional reason.
The differential equations from the coefficient of $T_{3}$ is  given by
\begin{eqnarray}
   \Big(\frac{d^2}{dr^2} + \frac{7}{r}\frac{d}{dr}\Big)f^{3}_0  +4f^3_2=0\,.
   \end{eqnarray}
Since $f^3_2=0$, nontrivial solution is of the form $f^{3}_0  \sim r^{-6}$, again not acceptable by dimensional reason.
 The remaining nontrivial part of differential equations  are 
 \beq
  \Big(\frac{d^2}{dr^2} + \frac{15}{r}\frac{d}{dr}\Big)f^{l}_4 =0\,, \qquad
   \Big(\frac{d^2}{dr^2} + \frac{11}{r}\frac{d}{dr}\Big)f^{l}_2 +16 f^{l}_4 =0\,,  \qquad   
   \Big(\frac{d^2}{dr^2} + \frac{7}{r}\frac{d}{dr}\Big)f^{l}_0 +4 f^{l}_2 =0\,,\eeq
where  $l =1,2$.  Relevant solutions for these equations are found as
\beq
f^{l}_4= \frac{c_l}{r^{14}}\,, \qquad 
f^{l}_2= -\frac{2}{3}\frac{c_l}{r^{12}}\,, \qquad 
f^{l}_0= \frac{1}{15}\frac{c_l}{r^{10}}\,, \qquad \qquad l =1,2\,,
\eeq
where $c_l$ is a numerical constant. After putting these solutions into the other equation (\ref{other}), we obtain $c_1 =c_2$. Therefore the eight fermion terms are completely determined up to an overall constant $c_1$, which should be a pure number. 
In~\cite{Baek:2008ws}, it was shown that non-zero $c_1$ comes from the one-loop corrections. Our results suggest that $|\p b|^4$ terms and their superpartners are one loop exact. It is interesting to note that for $k=1,2$ the supersymmetry of the effective action can be enhanced to $\CN=8$.

{\bf Acknowledgments}

We would like to thank the KIAS for hospitality.  S.H. was supported by
 the Science Research Center Program of the
Korea Science and Engineering Foundation through the Center for
Quantum Spacetime \textbf{(CQUeST)} of Sogang University with grant
number R11-2005-021.

\renewcommand{\theequation}{A-\arabic{equation}}
\setcounter{equation}{0}  
\section*{Appendix: Summary of useful properties}  
The $SO(2,1)$ gamma matrices, $\gamma_\mu$, satisfying $ \{\gamma_\mu,\gamma_\nu\} = 2\eta_{\mu\nu} $, obey Fierz identity 
\[  {(\gamma^{\mu})}_{\alpha}^{\, \beta}{(\gamma_{\mu})}_{\sigma}^{\, \delta}=2\delta_{\alpha}^{~\delta}\delta_{\sigma}^{~\beta}-\delta_{\alpha}^{~\beta}\delta_{\sigma}^{~\delta}\,. \]
The $SO(6)$ gamma matrices, $\Gamma_I$, satisfying 
$ \{\Gamma_I,\Gamma_J\} = -2\delta_{IJ} $,
may be represented as 
\bea\label{Gamma} \Gamma_I = \left(\ba{cc} 0 & \gamma_I \\ \bar{\gamma}_{I} & 0 \ea\right)\,, \qquad  \gamma^{\dagger}_I = -\bar{\gamma}_I\,,\eea
in the Weyl representation with
$  \bar{\Gamma}_{7} \equiv i\Gamma^1\Gamma^2 \Gamma^3 \Gamma^4 \Gamma^5 \Gamma^6= \diag ({\bf 1},  -{\bf 1} )\,. $
$\gamma^I$ are antisymmetric, $ \gamma^{~ AB}_{I} = - \gamma^{~ BA}_{I}$ and  related to $\bar{\gamma}^I$ as 
$
\gamma^{~ AB}_{I} = -\half \epsilon^{ABCD}\bar{\gamma}_{I\, CD}$, and satisfy
\beq
\gamma^{~ AB}_{I}\gamma^{I\, CD} = -2\epsilon^{ABCD}\,,   \quad   \bar{\gamma}_{I\, AB}\bar{\gamma}^{I}_{~ CD} = -2\epsilon_{ABCD}\,,  \quad
\gamma^{~ AB}_I\bar{\gamma}^{I}_{~ CD} = 2 (\delta^{A}_{~ C}\delta^{B}_{~ D}-\delta^{A}_{~ D}\delta^{B}_{~ C})\,.~~~  \label{gammarel} \eeq
Antisymmetric product of $\gamma$-matrices may be introduced as
\bea  \Gamma_{IJ} \equiv  \half \Big( \Gamma_I\Gamma_J - \Gamma_J\Gamma_I\Big) = \left(\ba{cc}  \gamma_{IJ} & 0\\ 0 & \bar{\gamma}_{IJ} \ea\right)\,, \qquad  (\gamma_{IJ})^{T} &=& -\bar{\gamma}_{IJ}\,.\eea
They are anti-hermitian,  $ \gamma^{\dagger}_{IJ} = - \gamma_{IJ}$,  $\bar{\gamma}^{\dagger}_{IJ} = - \bar{\gamma}_{IJ}$ and traceless, $\tr\, \gamma_{IJ} = \tr\, \bar{\gamma}_{IJ} =0$. They satisfy 
\bea \gamma^{~~~\, A}_{IJ~~ B}~ \gamma^{IJ~ C}_{~~~~~ D} &=& 2(\delta^{A}_{~ B}\delta^{C}_{~ D} - 4 \delta^{A}_{~D}\delta^{C}_{~ B})\,, \qquad
\bar{\gamma}^{~~~~~ B}_{IJ\, A}~ \bar{\gamma}^{IJ~~\, D}_{~~~ C} = 2(\delta^{~ B}_{A}\delta^{~ D}_{C} -4 \delta^{~\, D}_{A}\delta^{~\, B}_{C})\,, \nn\\ 
\gamma^{~~~ A}_{IJ~\, B}~  \bar{\gamma}^{IJ~~ D}_{~~~ C} &=& 2(4\delta^{A}_{~ C}\delta^{~ D}_{B} - \delta^{A}_{~ B}\delta^{~D}_{C})\,.   \nn
\eea
Similarly one can introduce
\bea  \Gamma_{IJK} =  \left(\ba{cc}  0  & \gamma_{IJK} \\ \bar{\gamma}_{IJK} & 0   \ea\right)\,, \qquad  \gamma^{\dagger}_{IJK} = \bar{\gamma}_{IJK}\,,\eea
where they are symmetric, $
 \gamma^{~~~~~ AB}_{IJK} =  \gamma^{~~~~~ BA}_{IJK}$.
They satisfy duality relations,
\[ \gamma_{IJK} = -\frac{i}{3!}\epsilon_{IJKLMN}\, \gamma^{LMN}\,, \qquad \bar{\gamma}_{IJK} = \frac{i}{3!}\epsilon_{IJKLMN}\, \bar{\gamma}^{LMN}\,. \]
as well as 
\[ \gamma^{~~~ AB}_{IJK}~  \bar{\gamma}^{IJK}_{~~~ CD} = 24(\delta^{A}_{~ C}\delta^{B}_{~D} + \delta^{A}_{~ D}\delta^{B}_{~ C})\,, \qquad
\gamma^{~~~ AB}_{IJK}~ \gamma^{IJK~ CD} =\bar{\gamma}_{IJK~ AB}~ \bar{\gamma}^{IJK}_{~~~ CD} =  0\,. \]
%


\thebibliography{999}
\bibitem{Aharony:2008ug}
  O.~Aharony, O.~Bergman, D.~L.~Jafferis and J.~Maldacena,
  [arXiv:0806.1218 [hep-th]].

\bibitem{Verlinde:2008di}
  H.~Verlinde,
  arXiv:0807.2121 [hep-th].

\bibitem{Baek:2008ws}
  J.~H.~Baek, S.~Hyun, W.~Jang and S.~H.~Yi,
  arXiv:0812.1772 [hep-th].

\bibitem{Berenstein:2008dc}
  D.~Berenstein and D.~Trancanelli,
  arXiv:0808.2503 [hep-th].
\bibitem{Hosomichi:2008ip}
  K.~Hosomichi, K.~M.~Lee, S.~Lee, S.~Lee, J.~Park and P.~Yi,
  JHEP {\bf 0811} (2008) 058
  [arXiv:0809.1771 [hep-th]].


  
  
  
\bibitem{Dine:1997nq}
  M.~Dine and N.~Seiberg,
  Phys.\ Lett.\  B {\bf 409} (1997) 239
  [arXiv:hep-th/9705057].
  
  
\bibitem{Paban:1998ea}
  S.~Paban, S.~Sethi and M.~Stern,
  Nucl.\ Phys.\  B {\bf 534} (1998) 137
  [arXiv:hep-th/9805018].
  
  
\bibitem{Paban:1998qy}
  S.~Paban, S.~Sethi and M.~Stern,
  JHEP {\bf 9806} (1998) 012
  [arXiv:hep-th/9806028].
  
\bibitem{Paban:1998mp}
  S.~Paban, S.~Sethi and M.~Stern,
  Adv.\ Theor.\ Math.\ Phys.\  {\bf 3} (1999) 343
  [arXiv:hep-th/9808119].

\bibitem{Hyun:1999hf}
  S.~Hyun, Y.~Kiem and H.~Shin,
  Nucl.\ Phys.\  B {\bf 558}, 349 (1999)
  [arXiv:hep-th/9903022].
\end{document}